\documentclass[conference]{IEEEtran}

\IEEEoverridecommandlockouts

\usepackage{cite}
\usepackage{amsmath,amssymb,amsfonts}
\usepackage{algorithm}
\usepackage{algpseudocode}
\algrenewcommand\algorithmicindent{0pt}

\usepackage{graphicx}
\usepackage{textcomp}
\usepackage{hyperref}
\usepackage{xcolor}

\begin{document}

\title{Graph Representation-Based Model Poisoning on the Heterogeneous Internet of Agents}

\author{
    \IEEEauthorblockN{Hanlin Cai$^{1}$, Haofan Dong$^{1}$, Houtianfu Wang$^{1}$, Kai Li$^{1,2}$, Sai Zou$^{3}$, Ozgur B. Akan$^{1,4}$}
    \IEEEauthorblockA{$^1$Internet of Everything (IoE) Group, Department of Engineering, University of Cambridge, Cambridge, UK}
    \IEEEauthorblockA{$^2$CISTER Research Centre, Porto, Portugal}
    \IEEEauthorblockA{$^3$Guizhou University, Guiyang, China}
    \IEEEauthorblockA{$^4$Center for neXt-Generation Communications (CXC), Koç University, Istanbul, Turkey}
    \IEEEauthorblockA{Email: \{hc663, hd489, hw680, oba21\}@cam.ac.uk, kai@isep.ipp.pt, zousai@gzu.edu.cn}    
    \vspace{-10pt}
}

\maketitle

\begin{abstract}

Internet of Agents (IoA) envisions a unified, agent-centric paradigm where heterogeneous large language model (LLM) agents can interconnect and collaborate at scale. Within this paradigm, federated fine-tuning (FFT) serves as a key enabler that allows distributed LLM agents to co-train an intelligent global LLM without centralizing local datasets. However, the FFT-enabled IoA systems remain vulnerable to model poisoning attacks, where adversaries can upload malicious updates to the server to degrade the performance of the aggregated global LLM. This paper proposes a graph representation-based model poisoning (GRMP) attack, which exploits overheard benign updates to construct a feature correlation graph and employs a variational graph autoencoder to capture structural dependencies and generate malicious updates. A novel attack algorithm is developed based on augmented Lagrangian and subgradient descent methods to optimize malicious updates that preserve benign-like statistics while embedding adversarial objectives. Experimental results show that the proposed GRMP attack can substantially decrease accuracy across different LLM models while remaining statistically consistent with benign updates, thereby evading detection by existing defense mechanisms and underscoring a severe threat to the ambitious IoA paradigm.

\end{abstract}


% **************************************************************
% **************************************************************
% **************************************************************

\section{Introduction}

The Internet of Agents (IoA) is an agentic paradigm in which heterogeneous intelligent agents can discover one another, interconnect, and collaborate at scale \cite{wang2025large}. In particular, IoA envisions large language model (LLM) agents as the core building blocks for realizing general-purpose reasoning, planning, and tool-using capabilities in both virtual and physical environments, where agents continuously acquire knowledge updates from the global interactions \cite{cai2025graph}. The IoA paradigm is expected to enable a broad spectrum of applications, such as personalized digital assistants, content generation, and decision-making support, while also imposing a fundamental requirement for scalable and resilient distributed fine-tuning across a population of LLM agents \cite{wang2025security}.

Federated fine-tuning (FFT) enables multiple LLM agents to collaboratively adapt a shared pretrained model while keeping all training data local, thereby satisfying privacy and data-residency constraints \cite{wang2025federated}. In FFT, each agent fine-tunes the model on its private dataset and periodically uploads local model updates to a coordinating server, which aggregates these updates to form a new global LLM and broadcasts it back for the next round of local training. To make FFT practical for billion-parameter LLMs under limited computation and wireless bandwidth, low-rank adaptation (LoRA) \cite{hu2022lora} is widely adopted: the LLM backbone is frozen and only lightweight low-rank parameters inserted into selected layers are trained and communicated. By transmitting only LoRA updates instead of full-model updates, FFT can reduce communication overhead and accelerate convergence, especially in the presence of heterogeneous data across the IoA system.

Despite the privacy-preserving advantages of the FFT-enabled IoA system, model poisoning attacks remain a critical resilience threat~\cite{han2024fedsecurity}. Specifically, the attacker operates by generating and transmitting malicious updates during the training process to poison the aggregated global LLM. Unlike conventional data poisoning, such an attacker does not require access to local datasets; instead, the attacker can exploit the inherent openness of wireless communications and the distributed nature of the IoA by participating as a legitimate but malicious agent and injecting adversarial objectives \cite{li2024biasing}.

Many defense methods have been developed to mitigate model poisoning attacks. In this paper, we refer to a broad family of such methods as \textit{DiSim-defense mechanisms}, which measure Euclidean distances or cosine similarities among local updates and flags those that appear as statistical outliers \cite{wan2024data}. For example, Krum \cite{blanchard2017machine} scores each local update by its Euclidean-distance consistency with a set of nearest neighbors and uses the most consistent update to guide aggregation, thereby suppressing outliers. FoolsGold \cite{fung2020limitations} uses cosine similarity among local updates to identify suspicious behavior and down-weights the corresponding contributions during aggregation. However, prevailing DiSim-defense mechanisms fundamentally depend on the statistical geometric coherence of local updates; if an attacker can embed adversarial objectives while preserving such coherence, it can therefore evade defenses.

\begin{figure*}[!ht]
    \centering
    \includegraphics[width=0.82\linewidth]{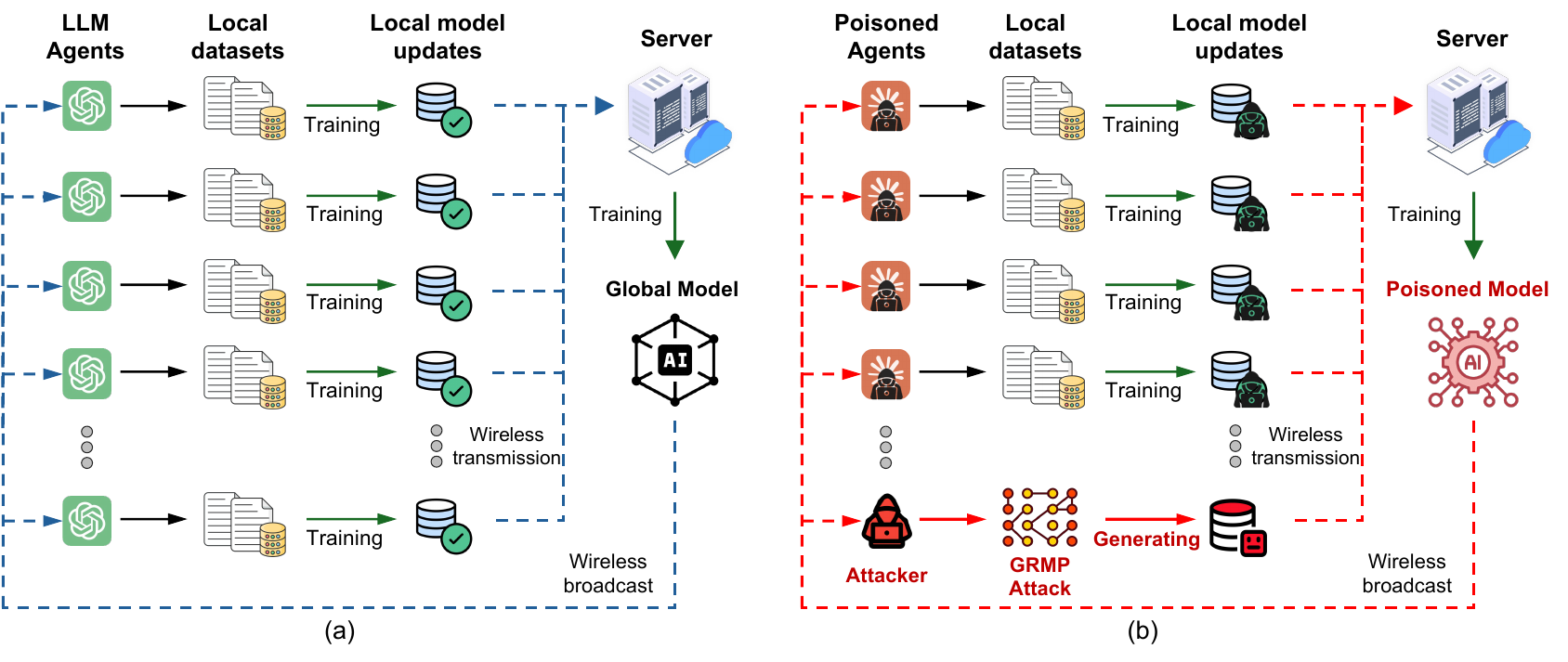}
    % \vspace{-12pt}
    \caption{(a) Training process of the IoA system, and (b) impact of the proposed GRMP attack on the IoA training cycle.}
    \vspace{-10pt}
    \label{fig:ioa}
\end{figure*}

In this paper, graph representation-based model poisoning (GRMP) is proposed as a novel attack method targeting the FFT-enabled IoA system. The GRMP attack uses graph representation learning to mimic the correlation features of benign updates, and employs a novel attack algorithm to optimize the malicious updates, thereby degrading the performance of the IoA system while bypassing DiSim-defense mechanisms.

\subsection{Our Contributions}

\begin{itemize}

    \item A novel GRMP attack is proposed to poison LLM agents within the IoA system. The GRMP attack leverages a variational graph autoencoder to generate malicious updates that preserve the statistical and geometric correlations of benign updates, thereby enabling the attacker to bypass existing DiSim-defense mechanisms.

    \item A new attack algorithm is designed based on the augmented Lagrangian and subgradient descent methods to iteratively optimize malicious updates, which can substantially influence the aggregated global LLM and cause a pronounced degradation in accuracy performance.

    \item The experiments are conducted with two LLM base models, DistilBERT and Pythia, and compare the proposed method against two existing attack methods. The results demonstrate that the proposed attack not only achieves substantially higher attack effectiveness, but also exhibits a level of statistical stealthiness that competing methods fail to attain. The GRMP is implemented based on PyTorch, and the source code has been released on GitHub: \href{https://github.com/GuangLun2000/IoA-Attack-GRMP}{https://github.com/GuangLun2000/IoA-Attack-GRMP}.
    
    % https://github.com/GuangLun2000/IoA-Attack-GRMP.

\end{itemize}

The remainder of this paper is organized as follows. Section \ref{sec2} describes the threat model and problem formulation. The proposed GRMP attack is presented in Section \ref{sec3}. Performance evaluation and resilience analysis are discussed in Section \ref{sec4}. Finally, Section \ref{sec5} concludes this paper.

% **************************************************************
% **************************************************************
% **************************************************************

\section{Problem Statement}\label{sec2}

\subsection{Federated Fine-Tuning (FFT) for LLM Agents}

As shown in Fig.~\ref{fig:ioa}(a), the IoA system comprises \(I\) benign LLM agents and no attackers. Each local agent \( i \in [1,I]\) maintains a private dataset \(\mathcal{D}_i\) of size \(D_i\) to train its local model, and the local datasets follow non-IID distributions across agents. Considering the billion-parameter scale of modern LLM and the constrained wireless bandwidth, a widely used solution is to perform FFT via LoRA \cite{hu2022lora}, where the backbone LLM is frozen and only the LoRA parameters are updated and transmitted. Let \( \mathbf w_i(t) \in \mathbb{R}^{1\times M}\) denote the vectorized LoRA parameters trained and communicated by agent \(i\) at communication round \(t\), where \(M\) is the LoRA parameter dimension. The local loss function of agent \(i\) in the \(t\)th communication round is
\begin{equation}
F\big(\mathbf w_i(t)\big)\!=\!\frac{1}{D_i}\!\sum_{(x,y)\in \mathcal{D}_i}\!\!f\Big(\mathcal{M}(x, \mathbf w_i(t)), y\Big),
\label{eq:local_obj}
% \!+\!\alpha\,\zeta\big(\mathbf w_i(t)\big),
\end{equation}
where $\mathcal{M}(x,\mathbf w_i(t))$ denotes the model output parameterized by $\mathbf w_i(t)$ on top of a frozen LLM backbone, and $f(\cdot,\cdot)$ is the task-specific loss function (e.g., cross-entropy) \cite{li2024data}. Upon completing local training in round $t$, each agent obtains $\mathbf w_i(t)$ and transmits its LoRA increment $\Delta\mathbf w_i(t)=\mathbf w_i(t)-\mathbf w_g(t-1)$ to the server, where $\mathbf w_g(t-1)$ is the global vectorized LoRA parameters broadcast at the beginning of round $t$. For notational simplicity, we refer to the local increment $\Delta\mathbf w_i(t)$ and the aggregated increment $\Delta\mathbf w_g(t)$ as the benign local update and the global update, respectively. The server aggregates the received benign updates as $\Delta\mathbf w_g(t)=\sum_{i=1}^{I}\frac{D_i}{\sum_{k=1}^{I}D_k}\Delta\mathbf w_i(t)$ and obtains the LoRA parameters by $\mathbf w_g(t)=\mathbf w_g(t-1)+\eta\,\Delta\mathbf w_g(t)$, where $\eta$ is the learning rate of the server.

% **************************************************************
% **************************************************************
% **************************************************************

\subsection{Threat Model}

In this paper, we propose a novel dataset-free model poisoning attack in which malicious updates are synthesized from overheard benign updates and correlation features extracted from benign and global updates. This attack can be particularly severe in wireless settings, owing to the broadcast nature of the radio medium \cite{cai2024securing}. As shown in Fig.~\ref{fig:ioa}(b), the attacker can impersonate a legitimate but malicious agent and passively overhear the local updates transmitted by some (if not all) benign agents, thereby generating and transmitting malicious updates \(\Delta\mathbf w'_j(t)\) to the server in each communication round. Unaware of the attacker’s presence, the server aggregates all received updates (benign and malicious) and obtains the contaminated global updates \(\Delta\mathbf w'_g(t)\). Thus, the contaminated global updates \(\Delta\mathbf w'_g(t)\) can be expressed as
\begin{equation}
\Delta\mathbf w'_g(t)=\sum_{i=1}^{I}\frac{D_i}{D}\Delta\mathbf w_i(t)+\sum_{j=1}^{J}\frac{D'_j}{D}\, \Delta \mathbf w'_j(t),
\label{eq:contaminated_global}
\end{equation}
where \(D\!=\!\sum_{i=1}^{I} \!D_i + \sum_{j=1}^{J} \!D_j'\), and \(D_j'\) is the data size claimed by attacker \(j\). The server then broadcasts the updated global LoRA parameters \(\mathbf w'_g(t)\) to all local agents as the reference for subsequent local training. To this end, the IoA system updates the global LLM agent using both benign and malicious local updates without accessing any local datasets.

The attacker’s goal is to exploit the correlations among the overheard benign updates \( \Delta\mathbf w_i(t)\) to synthesize an optimal \(\Delta\mathbf w'_j(t)\) that maximizes the global loss function \(F(\mathbf w'_g(t))\), while constraining the Euclidean distance between \(\Delta\mathbf w'_j(t)\) and \(\Delta\mathbf w'_g(t)\) within a reasonable range. This constraint enhances stealthiness, as the server could assess similarity metrics among local updates and filter out outliers that deviate substantially from the rest, e.g., Krum \cite{blanchard2017machine}. Thus, the optimization of the proposed GRMP attack launched by attacker \(j\) in the \(t\)th communication round can be formulated as
\begin{subequations}
\label{eq:attack_obj}
\begin{align}
\max_{\Delta\mathbf w'_j(t)} \quad 
& F\big(\mathbf w'_g(t)\big), \label{eq:attack_obj_a} \\
\text{s.t.}\quad
& d\big(\Delta\mathbf w'_j(t),\Delta\mathbf w'_g(t)\big)\ \! \le \! \ d_T(t), \label{eq:attack_obj_b}
\end{align}
\end{subequations}
where \(d(\cdot,\cdot)\) denotes the Euclidean distance, and \(d_T(t)\) is a dynamic threshold used to enforce stealthiness, set to the mean Euclidean distance (in round \(t\)) between the overheard benign local updates and the previous-round global update. The constraint \eqref{eq:attack_obj_b} ensures that the malicious local updates remain statistically consistent with the benign local updates.

% **************************************************************
% **************************************************************
% **************************************************************

\section{Proposed GRMP Attack}\label{sec3}

\subsection{Adversarial VGAE Model}

Due to the weak correlation between the parameter features of \( \Delta\mathbf w_i(t)\) and \( \Delta\mathbf w_j'(t)\), the server can detect malicious updates using DiSim-defense mechanisms. To address this, we design and train a new variational graph autoencoder (VGAE) to generate optimal \( \Delta\mathbf w_j'(t)\) that deeply captures the internal feature correlation of \( \Delta\mathbf w_i(t)\), thereby degrading the performance of the IoA system while bypassing such defenses.

We form the corresponding augmented Lagrangian function \cite{yue2025ug} of Problem \eqref{eq:attack_obj} as follows:
\begin{equation}
% \Delta\mathbf w'_j(t)
\label{eq:lagrangian}
\begin{aligned}
& \mathcal L \big(\Delta\mathbf w'_j(t),\lambda(t) \big)  =  F\big(\mathbf w'_g(t)\big)  -  \lambda(t)\big(d'(t) - d_T(t) \big) \\
& \ \ \ \ \ \ \ \ \ \ \ \ \ \ \ \ \ \ \ \ \ - \frac{\rho(t)}{2}\big(d'(t) - d_T(t) \big)^2,
\end{aligned}
\end{equation}
where \( d'(t) \! = \!  d\big(\Delta\mathbf w'_j(t),\!\Delta\mathbf w'_g(t)\big) \), \(\lambda(t) \geq 0 \) is the dual variable, and \(\rho(t) > 0\) is the penalty parameter. The Lagrangian dual problem for the constrained optimization in \eqref{eq:attack_obj} is
\begin{equation}
\label{eq:dual}
\left\{
\begin{aligned}
& \ \ \mathsf{D}(\lambda(t)) = \max_{\Delta\mathbf w'_j(t)} \ \mathcal L \big( \Delta\mathbf w'_j(t),\lambda(t) \big), \\[2pt]
& \ \min_{\lambda(t) \geq 0 }\ \, \mathsf{D}( \lambda(t)).
\end{aligned}
\right.
\end{equation}

To generate the optimal malicious updates \( \mathbf w_j'(t) \), we propose the VGAE model, which extends the variational autoencoder (VAE) framework \cite{cemgil2020autoencoding} to graph-structured data via unsupervised learning, aiming to maximize the augmented Lagrangian function in \eqref{eq:lagrangian} by solving
\begin{equation} \label{eq:wprime_sub}
    \Delta \mathbf w_j'(t)^\star = \arg\max_{\Delta \mathbf w'_j(t)}\ \mathcal L \big(\Delta\mathbf w'_j(t), \lambda(t) \big).
\end{equation}

Given \( \Delta \mathbf w'_j(t)^\star \), we can update \( \lambda(t) \) based on the subgradient descent method by solving the dual problem \ref{eq:dual}:
\begin{equation} \label{eq:update}
    \lambda(t+1)=\Big[\lambda(t)-\varepsilon\left(d'(t)-d_T\right)\Big]^{+}
\end{equation}
where $\varepsilon>0$ is the step size, and $[x]^+=\operatorname{max}\{0,x\}$.

\begin{figure*}[t]
    \centering
    \includegraphics[width=1.0\linewidth]{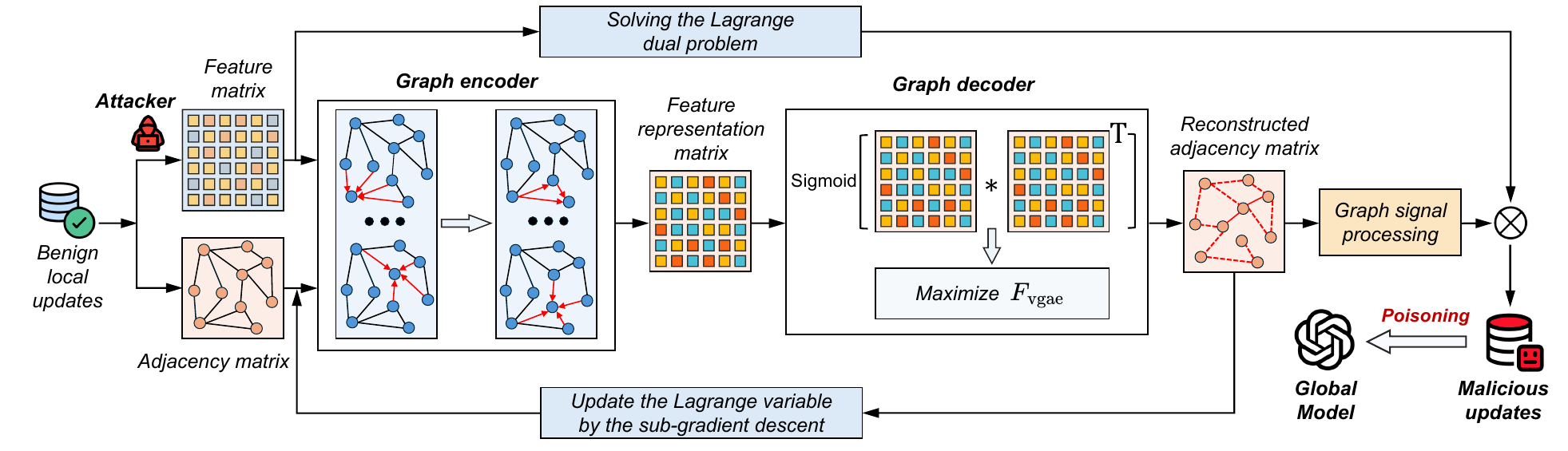}
    \vspace{-18pt}
    \caption{Framework of the proposed GRMP attack.}
    \vspace{-12pt}
    \label{fig:grmp}
\end{figure*}

As illustrated in Fig.\ref{fig:grmp}, the proposed GRMP attack iteratively optimizes the malicious updates \(\Delta\mathbf w'_j(t)\) based on the VGAE model and subgradient descent method. Using the observed benign updates $\Delta \mathbf w_i(t)$, the attacker estimates the internal correlation structures across LoRA parameters and encodes these relations as a graph $\mathcal{G}=(\mathcal{V},\mathcal{E},\mathcal{F})$, where the vertex set, edge set, and node feature matrix of the graph are represented by \(\mathcal{V}\), \(\mathcal{E}\), and \(\mathcal{F}\), respectively. The feature matrix \(\mathcal{F}(t) \! = \! [\Delta \mathbf{w}_{1}(t),\ldots, \Delta \mathbf{w}_{I}(t)]^{\mathsf T} \! \in \! \mathbb{R}^{I\times M}\) and the adjacency matrix \(\mathcal{A}(t) \! = \! [\delta_{m,m^{\prime}}(t)] \! \in \! \mathbb{R}^{M\times M}\) are the inputs to the VGAE model, where \(I\) is the number of observed benign updates, and \(M\) is the dimension of the benign updates. We use \(\delta_{m,m^{\prime}}(t)\) to denote the cosine similarity between \(w_m(t)\) and \(w_{m^{\prime}}(t)\), where \( w_m(t) \in  \mathbb{R}^{I\times 1} \) is the \(m\)th column of feature matrix \(\mathcal{F}(t)\), \(m,m^{\prime}\in[1,M]\), \( m \neq m^{\prime} \), and \(\delta_{m,m^{\prime}}\) is defined as
\begin{equation} \label{eq:cosine_similarity}
    \delta_{m,m^{\prime}}(t)=\frac{w_m(t)^{\mathsf T} \ w_{m^{\prime}}(t)}{\|w_m(t)\|_{2} \ \|w_{m^{\prime}}(t)\|_{2}}.
\end{equation}

Given \(\mathcal{F}(t)\) and \(\mathcal{A}(t)\), the topological structure of the graph \(\mathcal{G}\) can be constructed. The VGAE comprises a graph convolutional network (GCN) encoder and decoder. The encoder then maps \(\mathcal{G}\) into a lower-dimensional representation. We design the encoder based on a \(L\)-layer GCN architecture, which learns a representation that captures the internal features of \(\mathcal{G}\). The encoded representation is then fed into the decoder, which reconstructs the original graph using the lower-dimensional representation and generates the reconstructed adjacency matrix. Finally, a malicious local update \( \Delta \mathbf w_j'(t)\) is obtained based on graph signal processing and subgradient descent.

\subsubsection{Encoder of the VGAE} The encoder uses \(\mathcal{A}\) as input to its \(L\)-layer GCN. The output at the \(L\)-layer is defined as
\begin{equation}
\mathcal{Z}^L=f_\mathcal{G} \left(\mathcal{Z}^{L-1}, \mathcal{A} \mid \mathcal{W}^L \right),
\end{equation}
where \(f_\mathcal{G}(\cdot, \cdot\mid\cdot)\) is a spectral convolution function and \(\mathcal{W}^L\) is the weight matrix at the \(L\)-layer . Let \(\mathcal{I} \in \mathbb{R}^{M \times M}\) be the identity matrix in the GCN, we define \(\widetilde{\mathcal{A}}=\mathcal{A}+\mathcal{I}\) with the \((m,m')\)th matrix element \(\widetilde{\mathcal{A}}_{m,m'}\), and the diagonal degree matrix \(\widetilde{\mathcal{D}}\) with the \((m,m')\)th matrix element \(\widetilde{\mathcal{D}}_{m,m'} =\sum_{m^{\prime}=1}^M \widetilde{\mathcal{A}}_{ m, m^{\prime} } \). Thus, the encoder can be formulated as
\begin{equation}
f_{\mathcal{G}}\left({\mathcal { Z }}^{L-1}, \mathcal{A} \mid \mathcal{W}^L\right) = \phi \left(\widetilde{\mathcal{D}}^{-\frac{1}{2}} \widetilde{\mathcal{A}} \widetilde{\mathcal{D}}^{-\frac{1}{2}} {\mathcal { Z }}^{L-1} \mathcal{W}^L \right),
\end{equation}
where \(\phi(\cdot)\) is the activation function, e.g., ReLU\((\cdot)\).

\subsubsection{Decoder of the VGAE} The decoder aims to generate the original \(\mathcal{G}\) from its reduced representation. Let \( \widehat{\mathcal{A}} =\widehat{\mathcal{A}}(t)\) denote the reconstructed adjacency matrix, which is written as
\begin{equation}
\widehat{\mathcal{A}}(t)=\operatorname{Sigmoid}\left(\mathbf{Z}\left(\mathbf{Z}\right)^{\mathsf T}\right),
\end{equation}
where \(\operatorname{Sigmoid}(\cdot) \! = \! 1/(1 \! + \! \exp(-x))\) and \( \mathbf{Z} \! = \! \mathcal{Z}^L \). The larger inner product \(\mathbf{Z}(\mathbf{Z})^{\mathsf T}\) indicates a higher probability that the corresponding vertices \(\mathcal V_m\) and \(\mathcal V_{m'}\) are connected in \(\mathcal G\) \cite{li2024data}. The VGAE is trained by maximizing $F_{\mathrm{vgae}}$, which consists of a reconstruction term and a KL-regularization term, given by
\begin{equation}\label{eq:loss}
F_{\mathrm{vgae}} = \mathbb{E}_{q(\mathbf{Z} \mid \mathcal{F}, \mathcal{A})}[\log p(\mathcal{A} \! \mid \! \mathbf{Z})] \! - \! \operatorname{KL}(q(\mathbf{Z} \! \mid \! \mathcal{F}, \mathcal{A}) \| p(\mathbf{Z})),
\end{equation}
where \( \! p(\mathbf{Z}) \! \) is a Gaussian prior, \( \! \operatorname{KL}(q(\mathbf{Z} \! \mid \! \mathcal{F}, \mathcal{A}) \| p(\mathbf{Z})) \) is the Kullback-Leibler divergence between the variational posterior \(q(\mathbf{Z} \! \mid \! \mathcal{F}, \mathcal{A})\) and \( p(\mathbf{Z}) \), and the decoder likelihood \( p({\mathcal{A}} \! \mid \! \mathbf{Z}) \) indicates the correlation among the embedding vertices \cite{cemgil2020autoencoding}.

\subsubsection{Graph Signal Processing (GSP)}
As shown in Fig.~\ref{fig:grmp}, following the VGAE, the proposed attacker employs a GSP module to fuse the reconstructed matrices $\widehat{\mathcal{A}}$ and $\widehat{\mathcal{F}}$, thereby generating the malicious updates. The GSP module is designed to decompose the correlation features between different benign updates and the data features substantiating the updates.

A Laplacian matrix \(\mathcal L\) is constructed from the benign adjacency matrix \(\mathcal A\) as \(\mathcal L\!=\!\operatorname{diag}(\mathcal A \mathbf{1})\!-\!\mathcal A,\) where \(\operatorname{diag}(\cdot)\) maps a vector to a diagonal matrix. By applying singular value decomposition to \(\mathcal{L}\), e.g., \(\mathcal{L}\!=\!\mathcal{B} \Lambda \mathcal{B}^\top\), we can obtain an orthonormal matrix \(\mathcal{B}\in\mathbb{R}^{M \times M}\), also known as the graph Fourier transform (GFT) basis \cite{li2024biasing}, which is used to transform graph signals, e.g., \(\mathcal{F}\), to its spectral-domain representation. \(\Lambda\) is a diagonal matrix with the eigenvalues of \(\mathcal{L}\) on its diagonal.

Given \(\mathcal{B}\), the GRMP attacker projects the benign feature matrix onto the GFT basis to obtain the GFT coefficient matrix \(\mathcal{S}\! = \! \mathcal{F}\mathcal{B} \in \mathbb{R}^{I \times M}\), which captures the spectral-domain data features of observed benign updates. The attacker then constructs a reconstructed Laplacian matrix from the VGAE outputs as $\widehat{\mathcal L}=\operatorname{diag}(\widehat{\mathcal A}\mathbf 1)-\widehat{\mathcal A}$, and obtains the corresponding GFT basis $\widehat{\mathcal B}$ via eigen-decomposition of $\widehat{\mathcal L}$. Finally, the reconstructed feature matrix is recovered by $\widehat{\mathcal F}=\mathcal S\widehat{\mathcal B}^{\mathsf T}\in\mathbb R^{I\times M}$, where a row vector of $\widehat{\mathcal F}$ is selected as malicious local updates \( \Delta\mathbf w'_j(t) \) and transmitted to the server for aggregating the contaminated global update $\Delta \mathbf w'_g(t)$.

\subsection{Algorithm of the proposed GRMP Attack}\label{section:algorithm}
Algorithm~\ref{alg:grmp} outlines the workflow of the proposed GRMP attack, which is synchronized with the training process of the IoA system. Since the contaminated global LoRA parameters are broadcast by the server to all local agents, which then fine-tune their respective local LLM agents, the GRMP attack can propagate the malicious influence across the entire federation in an epidemic-like manner. Moreover, the GRMP attack is dataset-free: it operates by passively overhearing the IoA system's native wireless transmissions and does not require access local datasets of the benign agents. Consequently, the proposed GRMP attack incurs lower operational cost and risk than conventional data poisoning attacks \cite{wan2024data}, while remaining effective against existing DiSim-defense mechanisms.

\begin{algorithm}
\caption{Proposed GRMP attack algorithm}
\label{alg:grmp}
\begin{algorithmic}[1]
\State \textbf{Init:} \(\mathcal{G}(\mathcal{V},\mathcal{E},\mathcal{F})\), \(\eta\) ,\(T_l\), \( I \), \( J \). \( \lambda(1)\ge 0 \), and \( \rho(1) > 0\).
% \STATE \textbf{Init:} \(\lambda(1)\ge0\), \(\rho(1)\ge0\).
\For{round \(t=1,2,\ldots,T\)}
  \State Each benign agent \(i\) runs \(T_l\) local epochs to obtain \( \Delta \mathbf w_i(t)\); the attacker overhears \(  \Delta \mathbf w_i(t)\); the IoA server receives updates and aggregates the global update \( \Delta \mathbf w'_g(t)\).
  \State Attacker executes the VGAE and GSP to obtain \( \Delta \mathbf w_j'(t)\):
  \Statex \hspace{1.5em} $\bullet$ Calculate $\mathcal{A}$ according to \eqref{eq:cosine_similarity}, and input $\mathcal A$ and $\mathcal F$
    \Statex \hspace{2.5em} into the VGAE model.
  \Statex \hspace{1.5em} $\bullet$ Train the VGAE model to maximize \(F_{\mathrm{vgae}}\) by \eqref{eq:loss},
    \Statex \hspace{2.5em} and obtain the optimal $\widehat{\mathcal A}$.
  \Statex \hspace{1.5em} $\bullet$ Use the GSP module to obtain $\widehat{\mathcal F}$, and determine
    \Statex \hspace{2.5em} $ \Delta \mathbf w_j'(t)$ based on $\widehat{\mathcal F}$.
  \Statex \hspace{1.5em} $\bullet$ Update \(\lambda(t)\), according to \eqref{eq:update}.
  \State The attacker transmits $\Delta \mathbf w_j'(t)$ to the server.
  \State The server aggregates local updates to obtain the global LoRA parameters by \(\mathbf w'_g(t)=\mathbf w'_g(t-1)+\eta\,\Delta\mathbf w'_g(t)\).
  \State All local LLM agents update their model based on $\mathbf w_g'(t)$.
\EndFor
\end{algorithmic}
\end{algorithm}

% **************************************************************
% **************************************************************
% **************************************************************

\begin{figure*}[t]
    \centering
    \includegraphics[width=1.64\columnwidth]{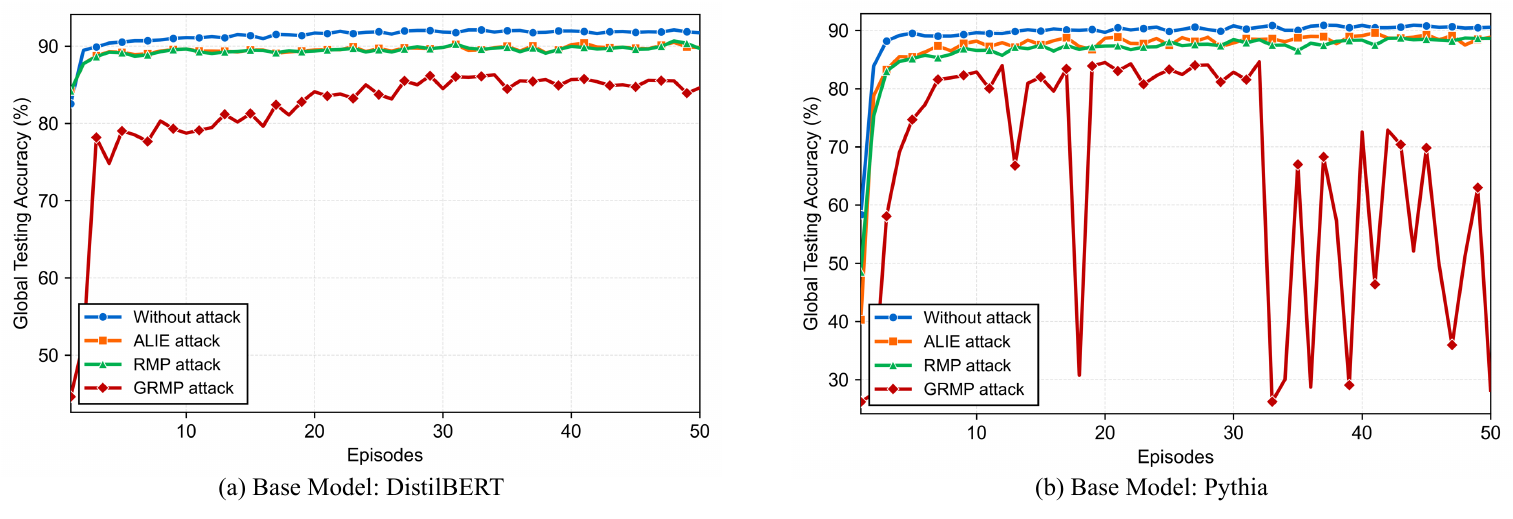}
    \vspace{-8pt}
    \caption{Global testing accuracy under the benign setting and under three attack methods over 50 communication rounds.}
    \vspace{-6pt}
    \label{fig:accuracy}
\end{figure*}

\begin{figure*}[t]
    \centering
    \includegraphics[width=1.88\columnwidth]{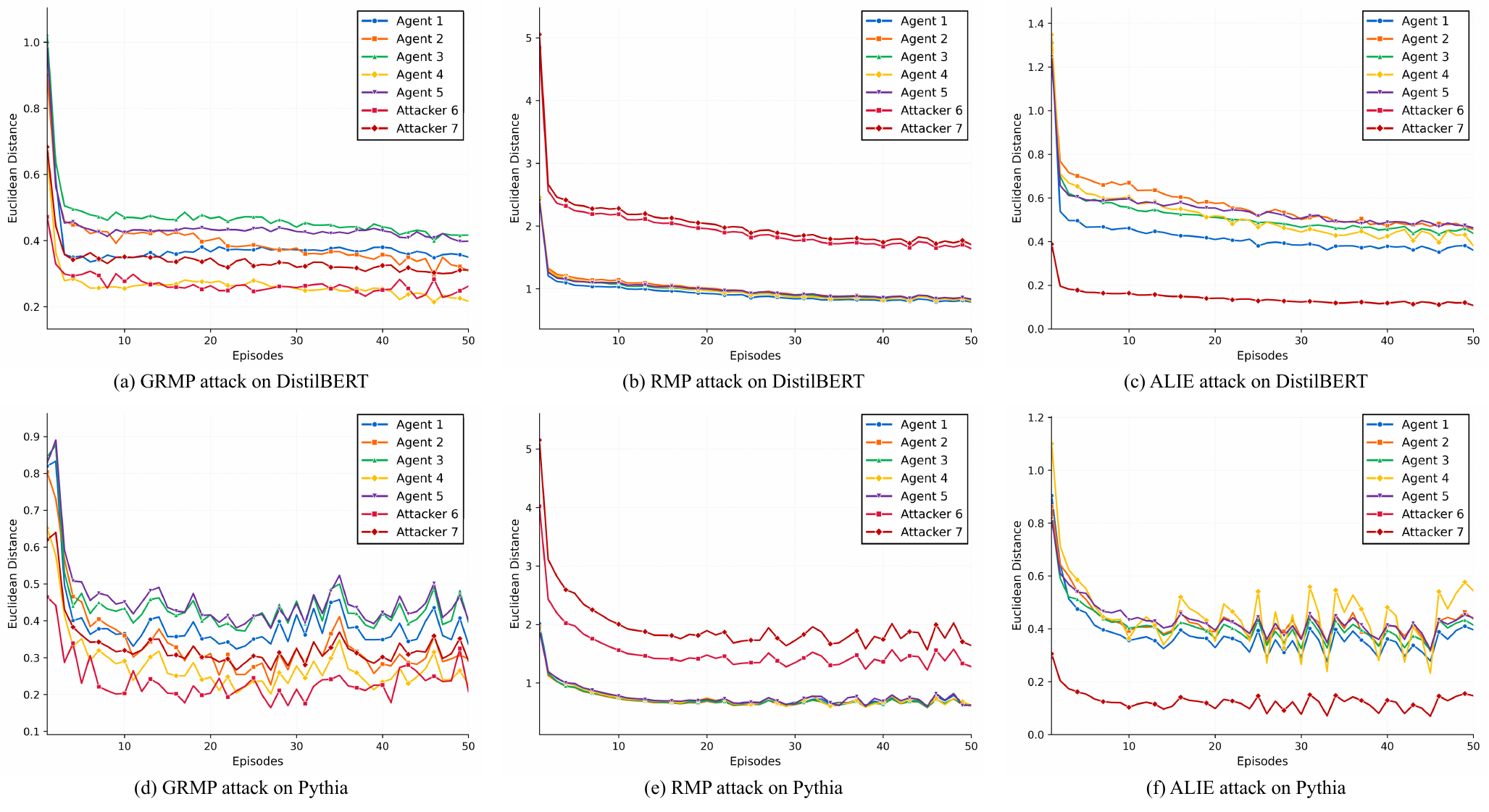}
    \vspace{-8pt}
    \caption{Euclidean distances between each agent's local update and the aggregated global update under three attack methods over 50 communication rounds.}
    \vspace{-8pt}
    \label{fig:distance}
\end{figure*}

\section{Performance Evaluation}\label{sec4}
    
To evaluate the effectiveness and stealthiness of the proposed GRMP attack, we conduct extensive experiments based on two LLM models and three attack methods. We use the AG News dataset \cite{zhang2015character}, a widely used benchmark consisting of four news categories (world, science, sports, and business), with 120,000 training samples and 7,600 test samples. We simulate an FFT-enabled IoA system with seven LLM agents, among which two are compromised by attackers. Training is performed for 50 communication rounds, where each agent runs five local epochs per round using DistilBERT (67M) \cite{sanh2019distilbert} or Pythia (160M) \cite{biderman2023pythia} as the pretrained base model. To simulate heterogeneous (non-iid) data distributions, we employ a Dirichlet distribution with the hyperparameter \( \mu  = 0.3 \). We adopt a standard LoRA configuration with rank $r = 8$, scaling factor $\alpha = 16$, and dropout rate $ p=0.1 $. 

The proposed GRMP attack is compared with two existing dataset-free baselines: ALIE attack \cite{baruch2019little} and the Gaussian random model poisoning (RMP) attack considered in \cite{cao2022mpaf}. Specifically, ALIE attack constructs malicious updates by shifting the mean of benign updates along the estimated standard-deviation direction, thereby producing statistically plausible yet adversarial perturbations. By contrast, RMP attack generates malicious updates by sampling from a Gaussian distribution estimated from benign updates and injecting these updates into the aggregation process as poisoning signals.

\subsection{Effectiveness Analysis}
Fig.~\ref{fig:accuracy} reports the testing accuracy of the global model under the benign setting and under three attack methods over 50 communication rounds. Fig.~\ref{fig:accuracy}(a) shows that, with DistilBERT as the base LLM, the accuracy of the benign global model quickly converges to approximately 92.0\%. ALIE and RMP attacks exhibit similar impacts, yielding average accuracies of 89.45\% and 89.39\%, respectively. The proposed GRMP attack is substantially more effective: it not only slows down convergence but also reduces the average accuracy to 81.69\%.

As shown in Fig.~\ref{fig:accuracy}(b), with Pythia as the base LLM, the benign global model achieves an average accuracy of 89.37\%, which is lower than that of DistilBERT. This performance gap is primarily attributable to differences in model design and pretraining objectives between Pythia and DistilBERT, which can lead to different levels of downstream sample efficiency under the same LoRA configurations. Under the ALIE and RMP attack, the average accuracies of the global model decrease to 86.84\% and 86.26\%, respectively. In comparison, the proposed GRMP attack substantially degrades the global model performance, causing the accuracy to drop sharply and then fluctuate dramatically between 25\% and 70\% thereafter.

\subsection{Stealthiness Analysis}

\begin{figure}[t]
    \centering
    \includegraphics[width=0.78\columnwidth]{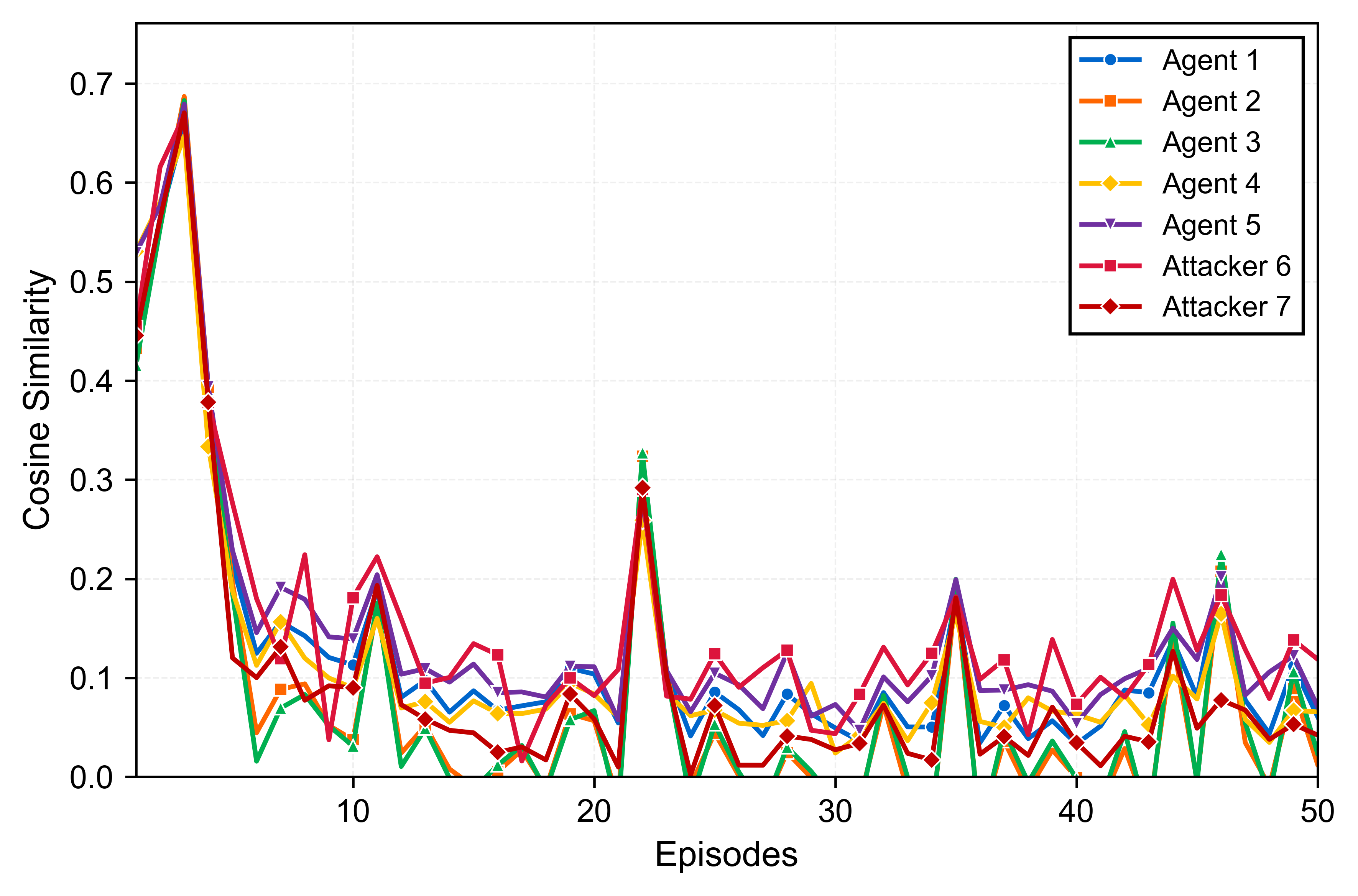}
    \vspace{-8pt}
    \caption{Cosine similarity of the local updates under the proposed GRMP attack over 50 communication rounds.}
    \vspace{-4pt}
    \label{fig:similarity}
\end{figure}

Fig.~\ref{fig:distance} illustrates the Euclidean distances between each local update and the aggregated global update under three attack methods over 50 communication rounds. A smaller Euclidean distance indicates that a local update is closer to the global update in parameter space. As shown in Fig.~\ref{fig:distance} (a) and (d), our proposed GRMP attack can capture the structural features of benign updates and synthesize malicious updates that closely follow benign parameter patterns while embedding adversarial objectives. By contrast, Fig.~\ref{fig:distance}(b)-(f) shows that the ALIE and RMP attacks fail to achieve such alignment: their Euclidean distances are either markedly larger or smaller than those of benign updates, causing the malicious updates to stand out as outliers and thus easier to detect. Moreover, Fig.~\ref{fig:similarity} further indicates that GRMP attack maintains a cosine similarity trajectory that nearly overlaps with the benign updates, reinforcing its stealthiness under both distance- and similarity-based defense metrics. These results underscore the key strength of the proposed attack: it can substantially degrade the performance of the IoA system while maintaining a benign-like update geometry, thereby evading existing DiSim-defense mechanisms.

% **************************************************************
% **************************************************************
% **************************************************************

\section{Conclusion}\label{sec5}

This paper proposes a novel dataset-free attack method, the GRMP attack, which exploits higher-order correlations among benign updates to synthesize statistically legitimate yet malicious local updates, thereby compromising the IoA system. Experimental results illustrate that the proposed GRMP attack can manipulate global aggregation and induce significant accuracy degradation while remaining elusive to existing defense methods based on Euclidean distance or cosine similarity. The GRMP attack raises a serious resilience concern for the ambitious IoA paradigm. Future work may explore adaptive aggregation and detection mechanisms that account for higher-order update dependencies, as well as principled defenses against correlation-preserving adversaries.

% **************************************************************
% **************************************************************
% **************************************************************

\section*{Acknowledgment}

This work is supported in part by the National Natural Science Foundation of China (62361011), National Research and Development Programs of China (2025YFB3109801),  Guizhou Provincial Science and Technology Plan Project (DXGA[2025]003, DXGA[2025]011), Guizhou Province Scientist Workstation (KXJZ[2025]005), Guizhou Province Science and Technology Innovation Platform (JSZX(2025)020), and China Scholarship Council (202508060002).

% **************************************************************
% **************************************************************
% **************************************************************

\bibliographystyle{IEEEtran}
\bibliography{ref}

\end{document}